\documentclass[aps,twocolumn,pre,showpacs,floatfix]{revtex4}
\usepackage{graphics,amssymb,amsmath}

\begin{document}

\title{Stochastic resonance in the driven Ising model on small-world networks}
\author{ H. \surname{Hong}}
\email{hhong@kias.re.kr}
\affiliation{Korea Institute for Advanced Study, Seoul 130-012, Korea}
\author{Beom Jun \surname{Kim}}
%\email{kim@tp.umu.se}
\affiliation{Department of Molecular Science
     and Technology, Ajou University, Suwon 442-749, Korea}
\author{M.Y. \surname{Choi}}
%\email{mychoi@phya.snu.ac.kr}
%\altaffiliation{Also at Korea Institute for Advanced Study, Seoul 130-012, Korea.}
\affiliation{Department of Physics, Seoul National University,
Seoul 151-747, Korea}

\begin{abstract}
We investigate the stochastic resonance phenomena in the field-driven Ising model
on small-world networks. 
The response of the magnetization to an oscillating magnetic field is examined 
by means of Monte Carlo dynamic simulations, with the rewiring probability varied.  
At any finite value of the rewiring probability, the system is found to undergo
a dynamic phase transition at a finite temperature, 
giving rise to double resonance peaks. 
%Here the shortcuts appear to play an opposite role in the ferromagnetic phase 
%and in the paramagnetic phase: 
While the peak in the ferromagnetic phase grows with the rewiring probability, 
that in the paramagnetic phase tends to reduce,
indicating opposite effects of the long-range interactions 
on the resonance in the two phases. 
%This interesting behavior is discussed in analogy to the opinion
%formation in a classroom with students and an instructor.
\end{abstract}
%05.45.Xt: Synchronization; coupled oscillators
%89.75.-k: Complex systems
%89.75.Fb: Structures and organization in complex systems
%05.40.-a: Fluctuation phenomena, random processes, noise, and Brownian motion
%75.10.Hk: Classical spin models
%89.75.Hc: Networks and genealogical trees

\maketitle

\section{Introduction}

It has been known that a periodically modulated bistable system may display
stochastic resonance (SR), arising from the cooperative interaction 
between random noise and periodic modulation~\cite{ref:SR:general}. 
Namely, when a system with an energetic activation barrier is subject to 
periodic but weak driving, the inherent thermal stochastic noise can 
enhance the signal out of the system rather than weaken it.  
Such SR phenomena, which have various practical 
applications~\cite{ref:SR:practical}, 
have been extensively investigated in a variety  of 
systems: In parallel to studies of systems with relatively small numbers
of degrees of freedom, the research focus has been shifting to
extended complex systems with many degrees of freedom, where
interesting collective dynamic behavior can 
emerge~\cite{ref:SR:complex,ref:ORfirst}. 
In those studies of SR in extended systems, the underlying connection
topology of dynamic variables has usually been assumed to be regular.
However, recent studies of computer networks, neuronal networks,
biochemical networks, and even social networks, have revealed that 
many real systems in nature possess quite complex structures,
which can be described neither by regular networks nor by
completely random networks~\cite{ref:network}. 

In this paper we consider an Ising model on Watts and Strogatz (WS) type 
small-world networks~\cite{ref:WS} and study the SR behavior 
in the presence of temporally oscillating external magnetic fields.
The WS network is characterized by a short characteristic path length
and high clustering~\cite{ref:WS}, both of which are commonly
observed properties of real networks in nature. 
Accordingly, we believe that the study of SR in an extended system
can be made more realistic if one uses the WS network 
as the underlying topology. 

There are five sections in this paper: Section II introduces the system,
i.e., the Ising model, driven by oscillating magnetic fields, on small-world
networks.  Defined in Sec. III is the dynamic order parameter, which describes 
conveniently the dynamic phase transition.  The relaxation behavior is 
examined and the relaxation time is computed.
Section IV is devoted to the stochastic resonance phenomena, 
characterized by double resonance peaks in the occupation ratio.  
The effects of long-range interactions on the resonance are
investigated, revealing opposite trends in the ferromagnetic and paramagnetic phases.
Finally, the main result is summarized and discussed in Sec. V.

\section{Ising Model on Small-World Networks}

The WS network in the present paper is constructed following
Ref.~\cite{ref:WS}: 
First, a one-dimensional regular network with only local connections (of range $k$)
is constructed with the periodic boundary condition.  
Next, each local link is visited once, and with the rewiring probability $P$ 
removed and reconnected to a randomly chosen node.  After the whole sweep of
the entire network, the average number of shortcuts in the network of 
size $N$ is given by $NPk$. 
Throughout this paper, the interaction range $k$ is set equal to two for convenience; 
longer ranges ($k >2$) are not expected to lead to any qualitative difference.
After the WS network is built as above, an Ising spin is put on every
node, and an edge (or a link) connecting two nodes is regarded as the coupling 
between the two spins at the two nodes.  

The Hamiltonian for the field-driven Ising model on the WS network thus reads
\begin{equation}
H = -\frac{J}{2}\sum_{i}\sum_{j\in\Lambda_i}\sigma_i \sigma_j 
-h(t)\sum_{i}\sigma_i,
\label{eq:Hamiltonian}
\end{equation}
where $J$ is the coupling strength, $\sigma_i \,(=\pm 1)$ is the Ising spin 
at node $i$, the neighborhood $\Lambda_i$ of $i$ stands for
the set of nodes connected to $i$ (via either local edges or shortcuts), 
and an oscillating magnetic field $h(t)=h_0 \cos\Omega t$ is applied
with the driving amplitude $h_0$ and frequency $\Omega$.  
We perform Monte Carlo (MC) dynamic simulations, 
employing the heat bath algorithm~\cite{ref:Newmanbook} and measuring
the time $t$ in units of the MC time step.
For thermalization, we start from sufficiently high temperatures and
lower the temperature $T$ slowly with the increment $\Delta T = 0.02$ 
(in units of $J/k_B$ with the Boltzmann constant $k_B$). 
The driving amplitude and frequency are mostly taken to be $h_0=0.1$ and $\Omega=0.1$ 
although different frequencies are also considered.  
While simulations are performed at a given temperature, 
the data from the first $4\times 10^4$ MC steps are discarded, 
which turns out to be sufficient for stationarity,
and measurements are made for next $10^4$ MC steps.
Networks of various sizes, up to $N=6400$, are constructed as described above, and 
averages are performed over 100 different network realizations.

In the absence of the long-range interaction ($P=0$), 
the network structure reduces to that of the one-dimensional
regular network with only local couplings.  
Accordingly, when $P=0$, the driven Ising model described by Eq.~(\ref{eq:Hamiltonian}) 
as well as the undriven model ($h_0 =0$) should not exhibit long-range order 
at finite temperatures.  
For $P \neq 0$, on the other hand, it has been found that the (undriven) Ising model 
displays ferromagnetic order at finite temperatures~\cite{ref:smallIsing}. 
This suggests that the driven Ising model in Eq.~(\ref{eq:Hamiltonian}) should
undergo a dynamic phase transition at a finite temperature
unless $h(t)$ is too large. 
When all nodes are fully connected, Eq.~(\ref{eq:Hamiltonian})
describes the infinite-range Ising model, where
double SR peaks have been observed and argued to be a generic property 
of the system with a continuous dynamic phase transition~\cite{ref:BJKim}.  
In view of this, similar double SR peaks are naturally expected when $P \neq 0$;
this contrasts with the case $P=0$, where the absence of a dynamic
phase transition implies the emergence
of just a single SR peak~\cite{ref:Brey}.  

\section{Dynamic Phase Transition}

\begin{figure}
\centering{\resizebox*{!}{5.5cm}{\includegraphics{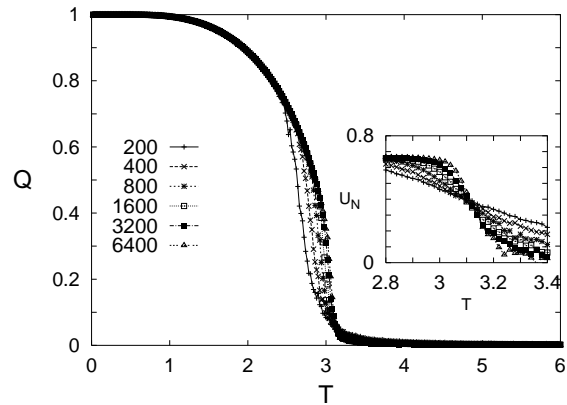}}}
\caption{Dynamic order parameter $Q$ versus the temperature $T$ in
the WS network of size $N=200$, 400, 800, 1600, 3200, and 6400,
at the rewiring probability $P=0.5$.
Inset: Dynamic fourth-order cumulant $U_N$ has a unique crossing point 
at $T_c\approx 3.13$ (in units of $J/k_B$).
}
\label{fig:Q}
\end{figure}

The dynamic phase transition in the system can be conveniently described by 
the dynamic order parameter. 
We first measure the magnetization 
\begin{equation}
m(t)\equiv \frac{1}{N}\sum_{i=1}^{N}\sigma_i , 
\end{equation}
and take the time average of $m(t)$ during the $n$-th period of $h(t)$,
to obtain 
\begin{equation}
Q_n \equiv \frac{\Omega}{2\pi}\int_{t_n}^{t_{n+1}} dt \,m(t)
\label{eq:Qn}
\end{equation}
with $t_n \equiv 2\pi n/\Omega$. 
The dynamic order parameter is then defined to be
\begin{equation}
Q \equiv \lim_{n \rightarrow \infty} Q_n , 
\label{eq:Q} 
\end{equation}
which takes different values
below and above the critical temperature $T_c$: 
Figure~\ref{fig:Q} shows that $Q\neq 0$ in the ferromagnetic phase at $T<T_c$
and $Q=0$ in the paramagnetic phase at $T>T_c$, 
with $T_c \approx 3$ for $P=0.5$. 
In order to determine $T_c$ more precisely, we measure 
the dynamic fourth-order cumulant~\cite{ref:Sides,ref:Binder}
\begin{equation}
U_N \equiv 1 - \frac{[\langle {Q_n}^4 \rangle]}{3[\langle {Q_n}^2 \rangle]^2},
\label{eq:U_N}
\end{equation}
where $\langle\cdots\rangle$ and $[\cdots]$ denote the time average 
and the average over different network realizations, respectively.
At $T=T_c$, the cumulant $U_N$ should have a unique value regardless of the
size of the system; this yields the estimation
$T_c \approx 3.13$, as shown in the inset of Fig.~\ref{fig:Q}.

\begin{figure}
\centering{\resizebox*{!}{5.5cm}{\includegraphics{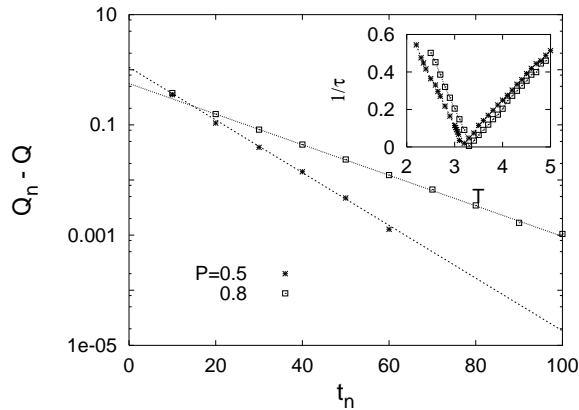}}}
\caption{Relaxation of the dynamic order parameter. 
$Q_n - Q$ versus $t_n$ at temperature $T=3.5$ is plotted in the semilog scale 
for the rewiring probability $P=0.5$ and 0.8. 
As the measurement time $t_n$ is increased, $Q_n$ is shown to approach 
its steady state value $Q$. 
Dashed lines represent the least-square fit to the exponential-decay form, 
$Q_n - Q \propto e^{-t_n/\tau}$,
from which one can obtain the relaxation time $\tau$. 
Inset: Inverse of the relaxation time $1/\tau$ versus the temperature $T$. 
} 
\label{fig:tau}
\end{figure}

To investigate the time relaxation behavior, 
we begin with the initial condition $m(t{=}0)=1$ 
(i.e., $\sigma_i = 1$ for all $i$), and measure $Q_n$ in Eq.~(\ref{eq:Qn})
as a function of $t_n$ during MC simulations. 
As time proceeds, $Q_n$ approaches the dynamic order parameter 
[see Eq.~(\ref{eq:Q})] with the value $Q=0$ and $Q \neq 0$ in the high- 
and low-temperature phases, respectively.
We find that the relaxation of $Q_n$ is very well described by the
exponential form, $Q_n - Q \propto e^{-t_n/\tau}$ with $t_n = 2\pi n /\Omega$,
which defines the relaxation time $\tau$~\cite{ref:relaxtime}. 
Figure~\ref{fig:tau} shows the relaxation behavior for $P=0.5$ and $0.8$
at $T=3.5$; it is observed that the exponential decay form describes
the simulation results very well.  We repeat the same procedure at other
temperatures and obtain the temperature dependence of the relaxation time,
shown in the inset of Fig.~\ref{fig:tau}.
As expected from the existence of the finite-temperature dynamic
phase transition, the relaxation time diverges near the dynamic
transition point ($T_c \approx 3.13$ and $3.20$ for $P=0.5$ and $0.8$,
respectively), and decreases as we move away from $T_c$ in both
directions~\cite{ref:BJKim}. 

\section{Stochastic Resonance}

In this section we study how the long-range interactions influence SR by varying
the rewiring probability $P$.  
The SR behavior is conveniently captured by the occupancy ratio $R$, 
first introduced in Ref.~\cite{ref:ORfirst} and
defined to be the average fraction of the spins in the 
direction of the external field~\cite{ref:BJKim:PRB}: 
\begin{equation}
R\equiv \left\langle\frac{\mbox{number of spins in the direction of}~h(t)}
{\mbox{total number of spins}}\right\rangle .
\label{eq:OR}
\end{equation}
In other words, $R$ measures how many spins follow the oscillating 
magnetic field. It is easy to understood
that $R$ has the value 1/2 in both low- and high-temperature limits
(see, e.g., Ref.~\cite{ref:BJKim}) and
becomes larger near the SR temperature, reflecting that more spins
follow the external driving. 

\begin{figure}
\centering{\resizebox*{!}{5.5cm}{\includegraphics{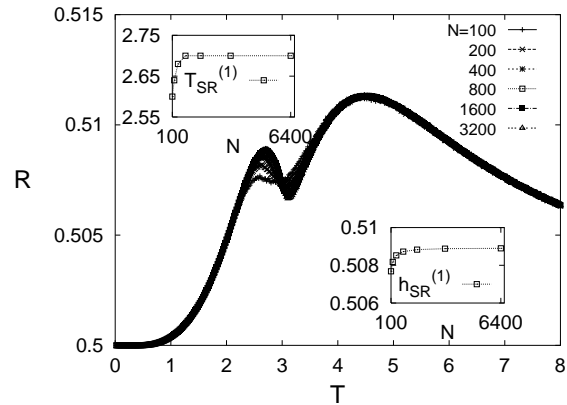}}}
\caption{Size effects of the occupancy ratio $R$ versus the temperature $T$ for  
$P=0.5$.  The finite-size effects are stronger for the SR peak
at the lower resonance temperature but become negligible for $N \gtrsim 800$.
Insets: Size-dependence of the SR temperature $T_{SR}^{(1)}$ and 
of the height $h_{SR}^{(1)}$ of the first resonance peak.}
\label{fig:size}
\end{figure}
In Fig.~\ref{fig:size}, $R$ in the system with $P=0.5$ is plotted 
as a function of the temperature for various sizes $N$. 
The double SR peaks are clearly exhibited, with the dip emerging in
the vicinity of the dynamic phase transition temperature $T_c$ 
determined from the crossing of $U_n$.
As the size $N$ is increased, the first SR peak at the lower resonance temperature 
$T_{SR}^{(1)}$ becomes sharper and apparently saturates for $N\gtrsim 800$,
displaying negligible finite-size effects.
The SR temperature $T_{SR}^{(1)}$ and the height $h_{SR}^{(1)}$ of the 
first resonance peak are displayed in the insets of Fig.~\ref{fig:size}, 
manifesting that there are no substantial size effects for $N \gtrsim 800$. 
For the second SR peak at the higher resonance temperature $T_{SR}^{(2)}$, 
finite-size effects are insignificant in both the height $h_{SR}^{(2)}$ and 
the SR temperature $T_{SR}^{(2)}$. 
We thus conclude that the double SR peaks are not merely finite-size effects,
and obtain $T_{SR}^{(1)}\approx 2.70$ and $T_{SR}^{(2)}\approx 4.46$ 
from the positions of the two SR peaks in Fig.~\ref{fig:size}. 

In Ref.~\cite{ref:BJKim}, the positions of SR peaks in the
infinite-range Ising model have been analytically obtained from 
the time-scaling matching condition:
The intrinsic time scale given by the relaxation time $\tau$ should
match the extrinsic time scale of the external driving.  
Furthermore, since $\tau$ diverges only at $T_c$ and decreases 
as the temperature $T$ is raised or lowered from $T_c$, the presence of double SR peaks 
should be a general property of the system with a finite-temperature
continuous phase transition. 
However, the specific form of the matching condition 
for the infinite-range Ising model, $\tau=1+\sqrt{1+\Omega^{-2}}$, 
found in Ref.~\cite{ref:BJKim} 
may not hold for the Ising model on the WS network studied in this work. 
Here we reveal the relation between the positions of 
double SR peaks and the relaxation time in the following manner: 
We first draw a vertical line at $T=T_{SR}^{(2)} \,(\approx 4.46)$ in 
the inset of Fig.~\ref{fig:tau} and 
locate the crossing point with the plot for $1/\tau$,
from which a horizontal line is drawn. 
This horizontal line corresponds to the time-scale matching condition 
and its crossing point with the lower temperature branch of the $1/\tau$ plot 
then yields the position of the first SR peak $T_{SR}^{(1)}$.  
This gives $T_{SR}^{(1)}\approx 2.70$, which is in excellent agreement with
the value obtained from the first peak of  $R$ in Fig.~\ref{fig:size}.

\begin{figure}
%\centering{\resizebox*{!}{5.5cm}{\includegraphics{fig4.eps}}}
\vspace*{16.0cm}
\includegraphics{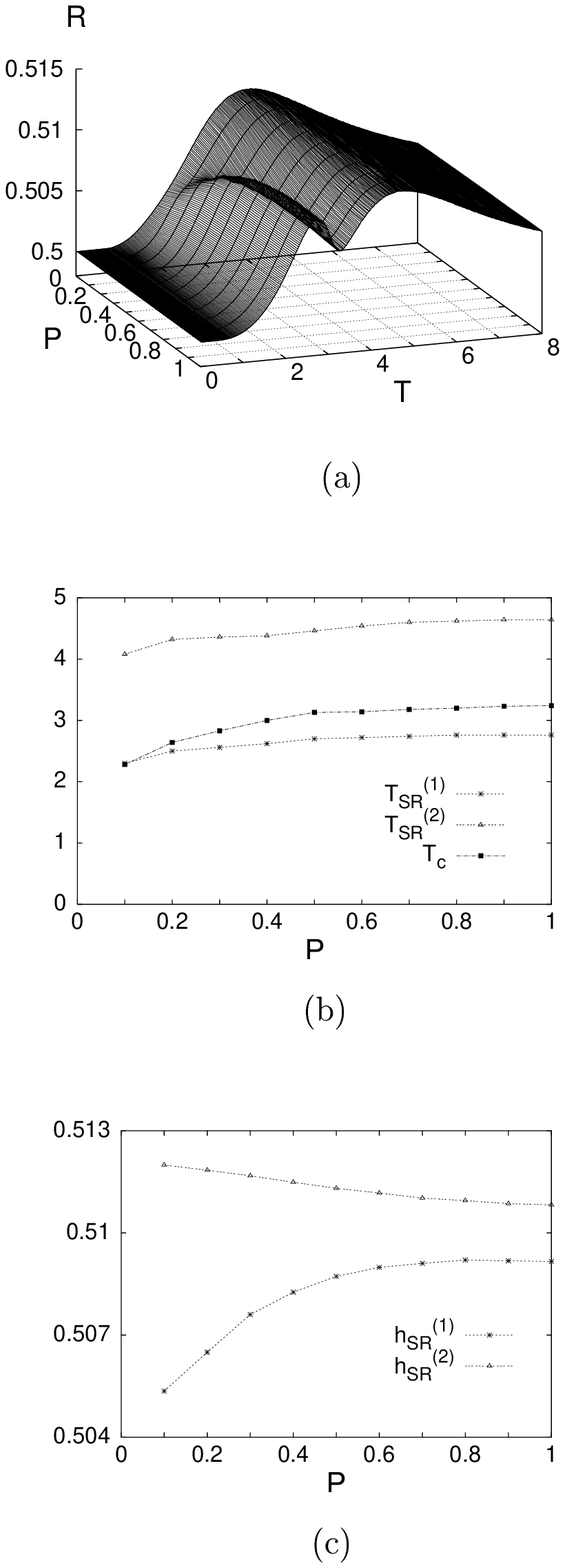}
\vspace*{1.2cm}
\caption{(a) Occupancy ratio $R$ versus the temperature $T$ and the rewiring 
probability $P$ in the system of size $N=800$. 
(b) Lower and upper resonance temperatures $T_{SR}^{(1)}$ and $T_{SR}^{(2)}$ 
as well as the dynamic transition temperature $T_c$ versus $P$. 
(c) Heights $h_{SR}^{(1)}$ and $h_{SR}^{(2)}$ of the resonance peaks versus $P$. 
}
\label{fig:OR}
\end{figure}
The behavior of $R$ with the temperature $T$ at various values of the rewiring 
probability $P$ is shown in Fig.~\ref{fig:OR}. 
When long-range interactions are not present ($P=0$), the Ising model
does not establish ferromagnetic order at finite temperatures, thus
not displaying a finite-temperature phase transition. 
Then expected is a single SR peak instead of two peaks~\cite{ref:BJKim},
which is confirmed in Fig.~\ref{fig:OR}. 
As $P$ is increased, the system behaves more similarly to the mean-field
system, and exhibits clear double SR peaks, 
which, in the mean-field (infinite-range) limit, have equal heights~\cite{ref:BJKim}.
Note that the maximally connected case ($P=1$) of the WS network has $O(N)$ 
long-range connections while the infinite-range system has $O(N^2)$ connections.  
This explains why the double peaks of $R$ at $P=1$ do not have equal heights
in Fig.~\ref{fig:OR}.

The SR peaks in Fig.~\ref{fig:OR} may be located from the condition 
$dR/dT = 0$ except for the first resonance peak at $P=0.1$, 
where we have used the condition that $dR/dT$ has a minimum positive value.  
In the insets of Fig.~\ref{fig:OR}, the obtained SR temperatures and peak heights 
are shown as functions of $P$.  It is noteworthy that as $P$ is increased the height
$h_{SR}^{(1)}$ of the first peak also grows, whereas $h_{SR}^{(2)}$ reduces. 
In other words, addition of long-range interactions tends to enhance SR in the
ferromagnetic phase, while the same addition suppresses SR in the paramagnetic phase. 
To understand these conflicting effects, which appear somewhat counterintuitive, 
we notice that as $P$ is increased, the critical temperature $T_c$ shifts toward higher
temperatures, reducing the relative (temperature) distance to 
$T_{SR}^{(2)}$ in comparison with that to $T_{SR}^{(1)}$. 
In the paramagnetic phase, the correlation length should increase with $P$ 
since $T_{SR}^{(2)}$ becomes closer to $T_c$; 
this leads more spins to be correlated and thus resistive to the change of 
the external driving field, resulting in the suppression of SR. 
In the ferromagnetic phase, on the other hand, a larger number of 
shortcuts (i.e., long-range connections) increases the distance to $T_c$, 
yielding a shorter correlation length. 
Accordingly, the spins are allowed to follow better the external driving,
and the SR is enhanced. 
It is also to be noted that SR in the present system tends to saturate for
$P\gtrsim 0.7$ and that almost the same SR as the 
random network ($P=1.0$) can be achieved with relatively small 
numbers of shortcuts.  Similar enhancing and saturating tendency has also been found 
in the synchronization behavior of the coupled oscillators on 
small-world networks~\cite{ref:Hong} as well as in the standard (single-peak) 
SR behavior of the system of coupled bistable elements~\cite{ref:SR:Hu}.

\begin{figure}
\centering{\resizebox*{!}{5.5cm}{\includegraphics{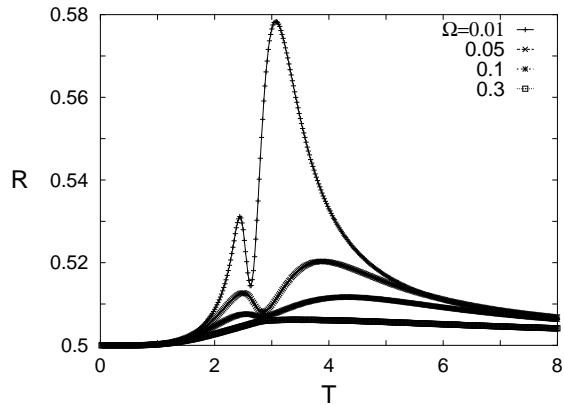}}}
\caption{Occupancy ratio $R$ versus the temperature $T$ for 
the rewiring probability $P=0.3$ and several driving frequencies. 
While the interval between the two peaks tends to grow with the driving frequency $\Omega$, 
the heights of the peaks in general reduce with $\Omega$. 
At high frequencies ($\Omega=0.3$), the lower peak in the ferromagnetic phase 
is observed to disappear.
}
\label{fig:diffW}
\end{figure}

We have also considered different driving frequencies and investigated how 
the occupancy ratio behaves.  Figure~\ref{fig:diffW} displays the behavior 
of $R$ depending on the temperature $T$ for various driving frequencies.
The interval between the two resonance peaks is observed to grow as the 
driving frequency is increased, which is consistent with the obtained relation 
between the relaxation time and the driving frequency.
On the other hand, the heights of the two peaks tend to reduce with the driving 
frequency; eventually, the lower peak in the ferromagnetic phase disappears, 
which reflects that at high driving frequencies 
there exists only one crossing point between the relaxation time 
and the driving frequency.
 
\section{Summary}

We have examined the stochastic resonance phenomena in the 
Ising model, driven by oscillating magnetic fields, on small-world networks.  
Double resonance peaks have been found to develop as the rewiring probability
is increased, and explained in terms of the diverging time scale 
at the dynamic phase transition.
It has been demonstrated that the resonance behavior essentially the same as 
that of the random network can be achieved with relatively small numbers of shortcuts.  
The most interesting finding in our system is that while stochastic resonance 
is enhanced with the number of long-range connections in the ferromagnetic 
phase, it is suppressed in the paramagnetic phase.  
As an interpretation based on analogy,
we suggest to consider the opinion formation in a social system:
an instructor and a group of students. 
More long-range connections in this analogy correspond to stronger interactions, 
i.e., more active discussions, between students far away, 
and thus help the group behave as a whole with a majority of students having
the same opinion (this situation is closely related to the
enhanced synchronization for larger $P$ in Ref.~\cite{ref:Hong}). 
However, this opinion developed from active
discussions among students may not necessarily be the opinion preferred by the
instructor of the class.  The enhanced SR peak in the ferromagnetic
phase corresponds to the situation that more interactions among students help the
class follow the instructor, with the students having already 
an identical opinion (whatever it is). 
On the other hand, the suppression of SR in
the paramagnetic phase is interpreted as the situation that 
more interactions make the students, among whom an agreement is not reached,
not so obedient to the instructor.

\acknowledgments
H.H. thanks J. Lee for providing the privilege of using the computing facility Gene.  B.J.K. acknowledges the support of Ajou Univ. for the year 2002.
M.Y.C. thanks the Korea Institute for Advanced Study for hospitality during his visit, 
where this work was performed, and
acknowledges the partial support from the Korea Research Foundation through
Grant No. 2000-015-DP0138.

\end{document}